\documentclass[fleqn,12pt,twoside]{article}
\usepackage{espcrc1}
\readRCS
$Id: espcrc1.tex,v 1.2 2004/02/24 11:22:11 spepping Exp $
\usepackage{graphicx}
\usepackage[figuresright]{rotating}

\newcommand{\AmS}{{\protect\the\textfont2
  A\kern-.1667em\lower.5ex\hbox{M}\kern-.125emS}}

\usepackage{amsmath}

\usepackage{epsfig}
\usepackage{graphicx}% Include figure files
\usepackage{dcolumn}% Align table columns on decimal point
\usepackage{bm}% bold math

\def\gapp{\lower.35em\hbox{$\stackrel{\textstyle>}{\sim}$}}
\def\lapp{\lower.35em\hbox{$\stackrel{\textstyle<}{\sim}$}}

\title{Interactions and disorder in 2D graphite sheets}

\author{Francisco Guinea\address{Instituto de Ciencia de Materiales de Madrid,
CSIC, Cantoblanco, E-28049 Madrid, Spain.},
M. Pilar L\'opez-Sancho\addressmark,
 and
 Mar\'{\i}a A. H. Vozmediano\address{Departamento de Matem\'aticas,
Unidad Asociada CSIC-UC3M,
\\ Universidad Carlos III de
Madrid, E-28911 Legan\'es, Madrid, Spain}}

\begin{document}

\maketitle
\section{Introduction}
Graphite has attracted a lot of recent attention due to the
growing evidence that, in many cases, it shows anomalous magnetic
and transport properties\footnote{See other contributions in this
volume.}.
 The understanding of these features is a significant
challenge. Theoretical models for the electronic
structure\cite{SW58} were developed under the assumption that
graphite could be described using the Landau theory of a Fermi
liquid, although it is assumed that the number of carriers is low.

The theoretical possibility of ferromagnetism in disordered
graphite samples was raised long ago\cite{OS91}. The underlying
mechanism is the existence of unpaired spins at defects, induced
by a change in the coordination of the carbon atoms (see below).
Experimental evidence showing  that the lifetimes of the
quasiparticles in graphite were not consistent with Fermi liquid
theory were reported in\cite{Yetal96}. A theoretical model
explaining these experiments was suggested in\cite{GGV96}. The
model used was based on the existence of incompletely screened
electron--electron interactions. In follow up work\cite{GGV01},
the analysis was extended in order to include the role of
disorder(see also\cite{SGH03}), which is known experimentally to
play an important role in relation to the existence of anomalous
magnetic properties.

The present work discusses theoretical models which address the
effects of electron--electron interactions and disorder in
graphene planes following the analysis in\cite{GGV01,SGH03}. The
starting point for the study is a simple tight-binding model for
the electronic structure, outlined in the next section. Then, a
discussion of the interesting features induced by the unscreened
Coulomb interaction is presented. The unusual features of the
model are emphasized. It is shown that the standard perturbative
treatments used in condensed matter theory fail, and a more
refined Renormalization Group approach is required. Then, a
theoretical framework which allows us to extend the model to many
types of disorder, following the approach in\cite{GGV92} for the
fivefold rings of the fullerenes, is discussed. The following
section analyzes the combined effects of disorder and
interactions. A brief discussion of models which may explain the
large anisotropy observed in very pure samples of
graphite\cite{Oetal03}, using the theoretical framework explained
in\cite{VLG02} is presented next. This work ends with a section
highlighting the most interesting conclusions.

We do not pretend to cover the large and rapidly growing
experimental literature on the magnetic properties of graphite and
related compounds. This work is extensively covered in other
chapters of this volume.

It is worth mentioning that the electronic structure of graphite
leads to theoretical models of significant interest for the
ongoing quest of understanding strongly correlated systems. This
work tries to underline also this aspect of the current work on
graphite and related compounds. Because of this reason, we also
include a brief summary of the technical aspects of the
calculations. We hope that this will not be discourage readers
willing just to grasp the main ideas of the work reported here.

\section{The elecronic structure of graphene sheets.}
\subsection{Description of the conduction band.}
Graphite, a three dimensional (3D) carbon-based material, presents
a layered and highly anisotropic structure, the interaction
between two adjacent layers being considerably smaller than the
intralayer interactions due to the large layer-layer separation,
3.35\AA \, when compared to the nearest-neighbor distance between
the carbon atoms $a = 1.42$\AA. In the planes, graphite exhibits
semimetallic behavior, and it presents a very weak electrical
conductivity along the perpendicular axis.

In the following, we consider the electronic structure of a single
graphite sheet, graphene.  In the 2D graphite the in-plane
$\sigma$ bonds are formed from 2$s$, 2$p_x$ and 2$p_y$ orbitals
hybridized in a $sp^2$ configuration, while the 2$p_z$ orbital,
perpendicular to the layer, builds up covalent bonds, similar to
the ones in the benzene molecule. The $\sigma$ bonds give rigidity
to the structure, while the $\pi$ bonds give rise to the valence
and conduction bands. The electronic properties around the Fermi
energy of a graphene sheet can be described by a tight binding
model with only one orbital per atom, the so-called $\pi$-electron
approximation, because, as stated above, the $\pi$ covalent bonds
are determinant for the electronic properties of graphite and
there are no significant mixing between states belonging to
$\sigma$ and $\pi$ bands in 2D graphite. Within this approximation
a basis set is provided by the Bloch functions made up of the
2$p_z$ orbitals from the two inequivalent carbon atoms A and B
which form the unit cell of the graphite hexagonal lattice.
Considering only nearest-neighbor interactions each atom A of a
sublattice has three nearest-neighbors B which belong to the other
sublattice \cite{Saito}.
\subsection{Tight-binding model.}
The nearest-neighbor tight binding approach reduces the problem to the
diagonalization of the one-electron Hamiltonian

\begin{align}
{\cal H} =  -t \sum_{<i,j>} a^+ _i a_j
\label{oneehamil}
\end{align}
where the sum is over pairs of nearest neighbors atoms $i,j$ on the lattice
and $a_i$, $a^+_j$ are canonically anticommuting operators

\begin{equation}
   \{a_i,a_j\} = \{a_i^+,a_j^+\} =0  ,    \{a_i,a_j^+\}=\delta_{ij}
\label{conmut}
\end{equation}

The eigenfunctions and eigenvalues of the Hamiltonian are obtained from
the equation
\begin{eqnarray}
\left(
\begin{array}{cc} \epsilon & -t\sum_j e^{ia{\bf k·u_j}}
\\ -t\sum_j e^{ia{\bf k·v_j}} &\epsilon \end{array} \right)
\left( \begin{array}{cc} C_A \\ C_B \end{array} \right)
= E( {\bf k }) \left( \begin{array}{cc} C_A \\ C_B \end{array} \right)\;,
\label{secuequt}
\end{eqnarray}
where ${\bf u_j}$ is a triad of vectors connecting an
A atom with its B nearest neighbors
and ${\bf v_j}$ the triad of their respective opposites,
$a$ is the distance between carbon
atoms and $\epsilon$ is the 2$p_z$ energy level,
taken as the origin of the energy.
The eigenfunctions, expanded as a linear combination of the atomic orbitals from
the two atoms forming the primitive cell, are determined by the the coefficients
$C_A$ and $C_B$ solutions of equation (\ref{secuequt}). The
eigenvalues of the equation  give the energy levels whose dispersion relation is

%\begin{figure}
%\centerline{\epsfig{file=.eps,width=5in}}
%\caption{Schematic dispersion relation around the Fermi points
%considered in the text.} \label{Fermi_points}
%\end{figure}

\begin{align}
E({\bf k})=\pm t\sqrt{1+4\cos^2\frac{\sqrt{3}}{2}ak_x+
4\cos\frac{\sqrt{3}}{2}ak_x\cos\frac{3}{2}ak_y} \;,
\label{disprel}
\end{align}
in which the two signs define two energy bands: the lower half called
the bonding $\pi$ band and the upper half called the
antibonding $\pi^*$ bands, which
are degenerate at the $ {\bf K}$  points of the Brillouin zone.
Within the $\pi$ electron approximation each site of the
graphite honeycomb lattice yields one electron
to the Fermi sea and the band is at half-filling.
Since each level of the band may accommodate two
states due to the spin degeneracy, and the Fermi
level turns out to be at the midpoint of the band,
instead a whole Fermi line,  the 2D honeycomb lattice has six isolated Fermi
points which are the six vertices of the hexagonal
Brillouin zone, two of which are inequivalent. The
lower branch of the dispersion relation is shown in Figure 1.
The calculation of the density of states shows that, at the Fermi level,
the density of states is zero therefore, the 2D graphite is a semiconductor of zero gap.
\begin{figure}
\centerline{\epsfig{file=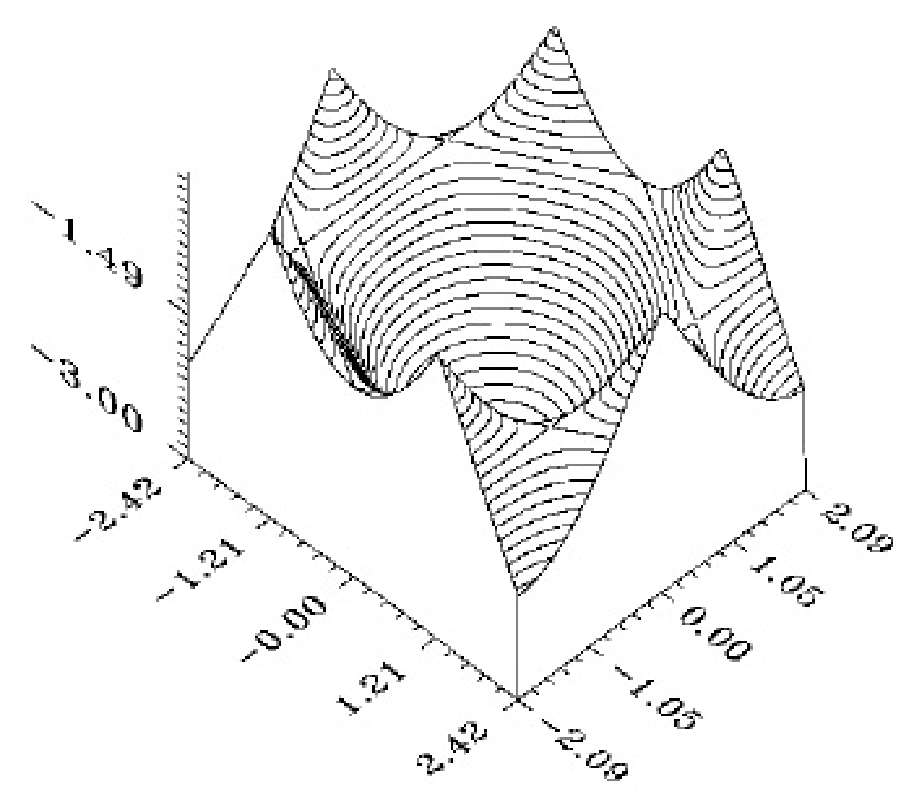,width=3in}} \caption{Lower
branch of the electronic dispersion relation. The cusps appear at
the six corners of the first Brillouin zone.} \label{fig1}
\end{figure}
The existence of a finite number of Fermi points at half-filling
has important consequences in the description of the spectrum
around the Fermi level. The low energy excitations can be studied
by taking the continuum limit at any two independent Fermi points.
As long as the number of the Fermi points is finite, the outcome
is that a simple field theory suffices to describe the electronic
spectrum of large honeycomb lattices. The continuum limit can be
taken by scaling of dimensionful quantities since we are dealing
with a free theory. Taken into account the parameter $a$, the
distance between carbon atoms, and expanding the 2x2 operator
(\ref{secuequt}) at any of two independent Fermi points, we have

\begin{eqnarray}
{\cal H}=
\left(
\begin{array}{cc} 0 & -t\sum_j e^{ia{\bf k·u_j}}
\\ -t\sum_j e^{ia{\bf k·v_j}} & 0 \end{array} \right)
\approx -\frac{3}{2}ta \left(
\begin{array}{cc} 0 &
\delta {\bf k_x}+\delta {\bf k_y} \\ \delta {\bf k_x}-
\delta {\bf k_y} & 0 \end{array} \right)
+O((a \delta {\bf k})^2)\;.
\label{equt}
\end{eqnarray}

The scaling
\begin{align}
\lim_{a\rightarrow 0} {\cal H}= -\frac{3}{2}t {\bf \sigma^T} \dot {\bf \delta {\bf k}}
\end{align}
determines the effective Hamiltonian in the continuum limit, which
turns out to be the Dirac operator in two dimensions. The same
result is obtained at any of the six ${\bf K}$ points of the
Brillouin zone, therefore, given the existence of the two
independent Fermi points, we conclude that the low energy
excitations of the honeycomb lattice at half filling are described
by an effective theory of the two-dimensional Dirac spinors. This
result is at odds with the more standard continuum approximation
to lattice theories in condensed matter physics, the effective
mass theory. In this theory, a quadratic dispersion relation at
high symmetry points of the Brillouin zone gives rise to an
effective Schr\"odinger equation, with one parameter, the mass,
chosen to reproduce the exact curvature. Only one dimensional
systems and three dimensional semiconductors with the diamond
structure and no gap, are known to give rise to the Dirac
equation.

\section{The long range Coulomb interactions in graphite.}
\subsection{Screening in graphite.}
The band structure of a graphene plane, as discussed in the
preceding section, leads to semimetallic behavior, as the density
of states vanishes at the Fermi energy. In a semimetal the long
range Coulomb interactions are not screened. The system, however,
has no gap, and we can expect that the electron--electron
interactions modify significantly the electronic structure near
the Fermi energy.

The role of the interactions  can be appreciated if their effect
is analyzed within perturbation theory. One obtains corrections to
the Fermi velocity and to the density of states which show a
logarithmic dependence on the temperature or other energy scale at
which these quantities are measured. This dependence implies that
perturbation theory cannot be used at sufficiently low energies.
On the other hand, it allows us to use the Renormalization Group
approach. In physical terms, the procedure amounts to defining
effective couplings which have a non trivial energy or temperature
dependence. The dependence of these couplings on energy can be
calculated using well tested techniques developed in the study of
Quantum Field Theory, as explained below.
\subsection {Renormalization group analysis of the interactions.}

The implementation of the renormalization group (RG)
scheme in condensed matter systems\cite{S94} has been a
theoretical hallmark for correlated electron
systems in the last decade.
The condensed matter approach shares
ideas from both the critical phenomena and the quantum field
theory approaches. The main issue is  that for special
systems (critical, renormalizable) the low-energy physics
is governed by an effective Hamiltonian made of
a few marginal interactions that can be obtained from
the microscopic high-energy Hamiltonian in a well prescribed manner.

Interactions  are classified as relevant, irrelevant or marginal
according to their scale dimensions. These dimensions determine
whether they grow, decrease, or acquire at most logarithmic
corrections at low energies. The effective coupling constants of a
model at intermediate energies by "integrating out" high-energy
modes even if there is no stable fixed point at the end of the RG
flow. The Luttinger and Fermi liquids are identified as infrared
fixed points of the RG applied to an interacting metallic system
in one or more dimensions respectively.

The main difficulty of the RG approach in condensed matter
systems in dimensions
greater than one lies on the
extended nature of the "vacuum" i.e., of the Fermi surface
what makes the issue of scaling rather tricky. The situation is
aggravated by the fact that the Fermi surface itself is changed
by the interactions, i.e. changes along the RG flow.

The Hamiltonian (\ref{hamil}) is the perfect model
for Renormalization Group (RG) calculations.
It is scale invariant and does not have the complications
of an extended Fermi surface. The model is similar to
the $D=1$ electron system\cite{d1}
in that it has Fermi points and linear
dispersion around them. Its two-dimensional nature manifests
itself in the fact that in this case four fermion interactions are
irrelevant instead of marginal.
The only interaction that may survive at low energies
is the long (infinite) range Coulomb interaction,
unscreened because of the vanishing density of states at
the Fermi point.

The RG analysis of the model is as follows:

The scaling dimension of the interactions are determined by these
of the fermion fields which can be read off from  the non
interacting hamiltonian,
\begin{align}
{\cal H}_{0}= \hbar v_F \int d^2 {\bf r} \bar{\Psi}({\bf r})
( i \sigma_x \partial_x + i \sigma_y \partial_y )
\Psi ({\bf r})
\end{align}
Because of the linear dispersion of the electronic states, we can
use $v_F$ to transform time scales into length scales. Then, we
can express the dimensions of all physical quantities in terms of
lengths. Within this convention, the Hamiltonian has dimensions of
energy ($l^{-1})$. This fixes the scale dimension of the
electronic fields to $[ \Psi ] = l^{-1}$, where $l$ defines a
length. This also ensures that the free Hamiltonian is scale
invariant. We can then readily determine the relevance of the
interactions to lowest order (tree level). The interacting
Hamiltonian including the two Fermi points $(i,i')$ and the spin
degrees of freedom $(s,s')$ is
\begin{eqnarray}
{\cal H}_{int} &= &\sum_{i,i';s,s'}
\frac{e^2}{2 \pi}
\int d^2 r_1 \int d^2 r_2 \frac{\bar{\Psi}_{i,s} ( \vec{r}_1 )
\Psi_{i,s} ( \vec{r}_1 ) \bar{\Psi}_{i',s'} ( \vec{r}_2 ) \Psi_{i',s'}
( \vec{r}_2 )}
{| \vec{r}_1 - \vec{r}_2 |} + \nonumber \\
&+ &\sum_{ s,s'; i,i'}  u_{i,s;i',s'}
\int d^2 r \bar{\Psi}_{i,s} ( \vec{r} )
\Psi_{i,s} ( \vec{r} ) \bar{\Psi}_{i',s'} ( \vec{r} )
\Psi_{i',s'} ( \vec{r} ) \;.
\label{hamil}
\end{eqnarray}
A naive power counting analysis shows that the Coulomb potential
(first term in eq. (\ref{hamil}))
defines a dimensionless, marginal coupling, while the four Fermi
couplings $u$'s
scale as $ l^{-1} $, and are irrelevant at low energies.
This effect can be traced back to the vanishing density of states
at the Fermi level.

The next step of the RG consists in analyzing the renormalization
of the parameters describing the system when quantum
corrections are included. When renormalized,
the marginal interactions can either
grow, driving the system away of its free fixed point -- this
is the case of an attractive four Fermi interaction in the
Fermi liquid case --, decrease and become irrelevant (repulsive
interactions in the Fermi liquid), or stay marginal in which case
they define the theory (forward scattering in a Fermi liquid
and the related Landau parameters). Our model differs from the
usual Fermi liquid analysis of \cite{S94} on the fact that
our interaction is a long ranged (infinite range) unscreened
Coulomb interaction, a case that lies away of the Fermi liquid hypothesis.

Following the quantum field theory nature of the model, we replace
the instantaneous Coulomb interaction of eq. (\ref{hamil})
\begin{align}
{\cal H}_C=
\frac{e^2}{4\pi v_F}
\int d^2 {\bf r_1} \int d^2 {\bf r_2} \frac{\bar{\Psi}({\bf r_1})
\Psi ({\bf r_1})  \bar{\Psi}({\bf r_2})  \Psi({\bf r_2})}
{|{\bf r}_1 - {\bf r}_2 |}
\end{align}
where $g=e^2/4\pi v_F$ is the dimensionless coupling constant,
by a local gauge interaction through a minimal coupling.
%\begin{align}
$$L_{int}=g\int d^2x dt j^\mu (x,t)A_\mu (x,t)\;, $$
%\end{align}
where the electron current is defined as
$$j^\mu =(\overline\Psi\gamma^0\Psi,v_F \overline\Psi\gamma^i\Psi)\;,$$
the three $\gamma$ matrices ($\gamma^{0,1,2}$) are appropriate
combinations of the Pauli matrices. The full Hamiltonian is then
that of (non-relativistic) quantum electrodynamics in two spacial
dimensions, a model used also in the physics of nodal states of
d-wave superconductors:
\begin{align}
{\cal H}= \hbar v_F \int d^2 {\bf r} \bar{\Psi}({\bf r})
\gamma^\mu(\partial_\mu -igA_\mu)
\Psi ({\bf r})\;.
\label{full}
\end{align}
The RG analysis proceeds with the computation of the
renormalization of the parameters of the model. The Feynman
diagrams building blocks are the free electron and photon
propagators:
$$G_0(\omega, {\bf k})=i\frac{-\gamma^0\omega+v_F
\gamma \cdot {\bf k}}{-\omega^2+v_F^2{\bf k}^2-i\epsilon}\;\;,\;\;
\Pi^0_{\mu\nu}(r_1, r_2)=-i\delta_{\mu\nu}\int \frac{d^4 k}{(2\pi)^4}
\frac{e^{-i\omega(t_1-t_2)}e^{i{\bf k}({\bf x_1}
-{\bf x}_2)}}{-\omega^2+{\bf k}^2-i\epsilon}\;.
$$

The electron self--energy $\Sigma(\omega, {\bf k})$ defined by the
equation $G^{-1}=G_0^{-1}-\Sigma \;$, is renormalized by the
Feynman diagrams of Fig. 2.
\begin{figure}
\centerline{\epsfig{file=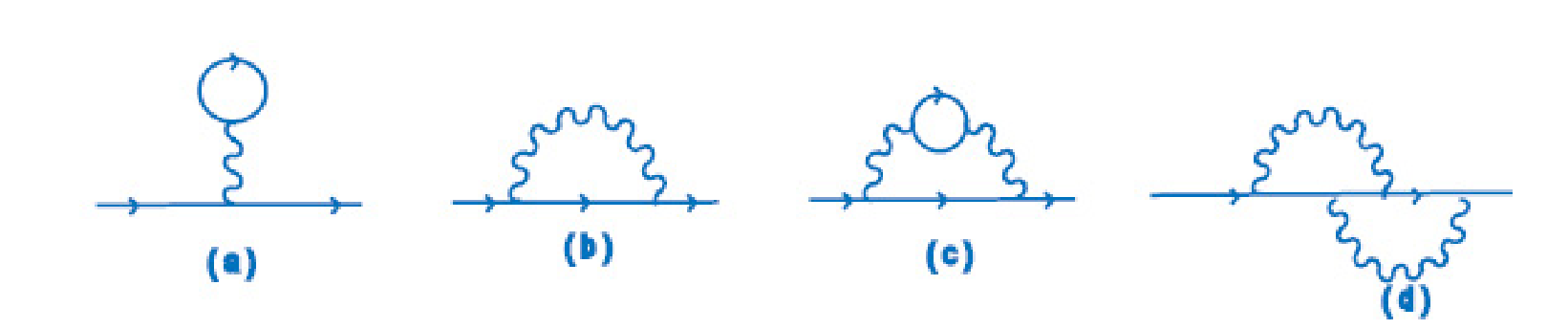,width=5in}} \caption{Feynman
diagrams renormalizing the electron self-energy.} \label{fig2}
\end{figure}
It contains the following physical information:
\begin{itemize}
\item The density of states $n(\omega)=Im \int d^2 {\bf k}\;{\rm
tr} G(\omega, {\bf k})\sigma_3\;.$ It is renormalized by the
diagram in Fig. 2$a)$. \item The Fermi velocity renormalization.
It is obtained already at the one loop level from the  diagram in
Fig. 2$b$). \item The quasiparticle lifetime $\tau^{-1}\sim
\lim_{\omega\to 0}{\rm Im }\Sigma(\omega, {\bf k})\;.$ Its first
contribution is at the two--loops level from diagrams 2$c$), 2$d$)
in Fig. 2. \item The wave function renormalization
$Z_\Psi\sim\frac{\partial \Sigma(\omega, {\bf
k})}{\partial\omega}\vert_{\omega=0}\;,$ defines the anomalous
dimension of the field $\gamma=\partial \log Z_\Psi/\partial l$
($l$ is the RG parameter) and, hence of the fermion propagator:
$G(\omega, {\bf k})\sim_{\omega\to 0}\frac{1}{\omega^\eta}\;.$ It
is a critical exponent that determines the universality class of
the given model. Under the physical point of view it affects the
interlayer tunneling and other transport properties.
\end{itemize}

The next set of diagrams to analyze corresponds to the
photon self--energy and vertex corrections
represented schematically in Fig. 3.
\begin{center}
\begin{figure}
\centerline{\epsfig{file=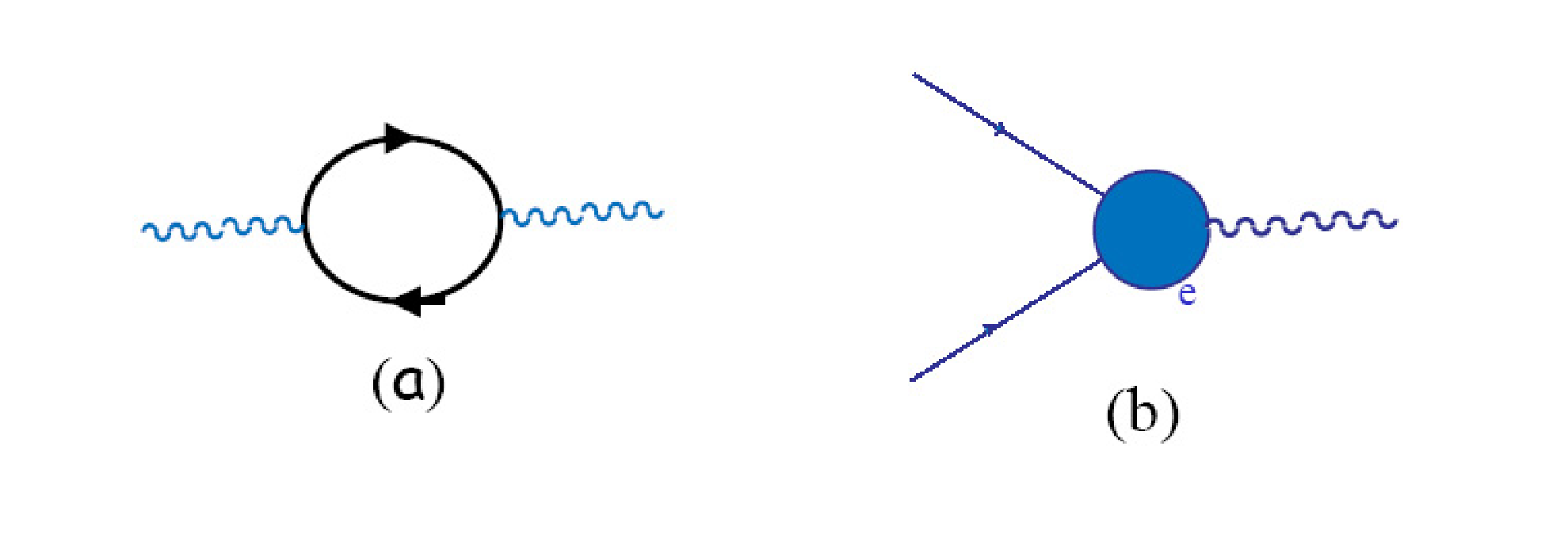,width=4in}}
\caption{($a$) Feynman diagram renormalizing the photon
self-energy. (b) Vertex correction.} \label{fig3}
\end{figure}
\end{center}
The real part of the photon self-energy
at one loop (polarization) renormalizes the
interaction and the imaginary part gives the density of
electron-hole excitations of the system. The vertex corrections
renormalize the electric charge.
\subsection{Results.}
In the computation of the diagrams mentioned it is readily seen
that the loop corrections come in powers of an effective
coupling constant given by $g=e^2/4\pi v_F$. The physical
results extracted from the RG analysis are the following.

1. From the computation of the electron self-energy
(Fig.   2$b$)) we get a non trivial renormalization
of the Fermi velocity that {\bf grows in the infrared}. This
result implies a breakdown of the relation between the energy
and momentum scaling, a signature of a quantum critical point.

2. From the electron self-energy at two loops order we get a non
trivial wave function renormalization meaning that the infrared
stable fixed point corresponds to a free fixed point different
from the Fermi liquid. This result has been shown to persist in
the  non-perturbative regime\cite{GGV99}. This is a non-trivial
result that has physical implications. In particular it implies
that the inverse quasiparticle lifetime increases linearly with
energy\cite{GGV96}, a result that has been observed experimentally
in \cite{Yetal96} in the energy range of validity of the model.

3. The electron-photon vertex and the photon propagator are not
renormalized at the one loop level. This means that the electric
charge is not renormalized, a result that could be predicted by
gauge invariance, and it also implies that the effective coupling
constant $g=e^2/4\pi v_F$ {\bf decreases at low energies} defining
an infrared free fixed point of the RG. It is interesting to note
that the Lorentz invariance of the model that was explicitly
broken by the Fermi velocity is recovered at the fixed point since
the velocity  of light, c, fixes a limit to the growing of the
Fermi velocity. In conclusion, the RG analysis shows that without
disorder, edges, or other perturbations, the graphene system at
low energies has gapless excitations differing from the Fermi
liquid quasiparticles but does not support magnetic or
superconducting instabilities. It is interesting to note that the
energy dependence of the coupling constant\cite{GGV99} can lead to
non trivial scaling features in optical properties\cite{KM04}.

The strong coupling regime of the graphene system has been
analyzed in \cite{Khv}. There it is argued that a dynamical
breakdown of the chiral symmetry (degeneracy between the two Fermi
points) will occur at strong coupling and a gap will open in the
spectrum forming a kind of charge density wave. Graphite can then
be seen as an excitonic insulator that can become ferromagnetic
upon doping. The resulting gap has an exponentially small
non-perturbative value.

The analysis in this section neglects short range interactions, as
their effects are less relevant than those arising from the long
range Coulomb interaction. It is worth noting, however, that a
sufficiently large on--site repulsion can induce a transition to
an antiferromagnetic ground state\cite{ST92}, and that, even below
this transition, significant effects at low energies can be
expected\cite{BJ02}.
\section{Effects of disorder.}
\subsection{General features.} As mentioned elsewhere in this
volume, there is a wide variety of carbon compounds, ranging from
crystalline diamond, where the carbon atoms show fourfold
coordination, to graphite, where the coordination is threefold,
and the coupling between neighboring planes is weak. The
environment around a carbon atom in nanotubes and the fullerenes
is closer to the graphite case, although the bonds with the three
nearest neighbors are distorted. The variety of possible
environments around a carbon atom imply that many intermediate,
metastable phases can exist. As in any other materials, disorder
can appear due to lattice defects or impurities. In the following,
we consider the changes in the electronic states in threefold
coordinated systems due to some simple lattice defects, like five-
and sevenfold rings, vacancies, dislocations and edges. We will
not address the stronger deformations associated with hybrid
three-and fourfold bonding ($sp^3-sp^2$ hybridization) which may
exist in highly disordered systems\cite{OS91}.

%\section{Electronic properties of extended defects.}
\subsection{Five- and sevenfold rings (disclinations).} As
discussed earlier, the low energy electronic states of graphene
planes are well described by a two dimensional Dirac equation,
which reproduces correctly the semimetallic nature of the system.
Some lattice distortions give rise to long range modifications in
the electronic wavefunctions. These effects should be well
described using the effective Dirac equation as a starting point.

The simplest defects which show these features are five- or
sevenfold rings in the honeycomb lattice. These defects can be
considered disclinations of the lattice, which acquires a finite
curvature. The accumulation of them leads to curved shapes, like
the fullerenes, which show twelve fivefold rings. Sevenfold rings
lead to negative curvature, and a variety of compounds have been
proposed to exist with this property\cite{VT92}. A simple way to
show that an odd numbered ring in the honeycomb lattice leads to
long range effects in the electronic spectrum is by noting that
any closed path which encompasses the defect leads to an
interchange of the two sublattices which build the
structure\cite{GGV92}. The description of the electronic states in
terms of the Dirac equation is achieved by using to types of
electronic ^^ ^^ flavors ", each of them existing in a different
sublattice. The existence of odd numbered rings changes the Dirac
equation at any distance from the defect.

If we neglect for the moment the effect of the long range lattice
distortions induced by these defects, the only consequence of the
presence of odd numbered rings is the above mentioned interchange
of the two sublattices. In the Dirac description it implies that,
when moving around the defect, the two electronic flavors are
exchanged. The standard way to associate to a translation a smooth
change in other properties is through gauge potentials. The
existence of a gauge potential implies, in general, that the usual
derivative has to be replaced by the covariant derivative, which
includes the potential. The usual derivative operator is the
generator of a translation through the system. A covariant
derivative with a finite gauge potential implies that, when
translating an object, an additional operation has to be performed
upon it. In the case considered here, this operation is a rotation
in flavor space. As there are two flavors, the index which
distinguishes them is equivalent to a spin one half. The rotations
in this space build up the SU(2) non abelian group. The gauge
potential needed has to be chosen such that the accumulated
rotation in a path which encircles the defect should be
independent of the path. Hence, the gauge potential is equivalent
to that generated by a fixed ^^ ^^ magnetic " flux at the position
of the defect\footnote{One must note that there is an additional
technical complication, associated to the fact that there is also
another index associated to the two inequivalent Fermi points in
the Brillouin Zone. An odd numbered ring also exchanges them.}.

\begin{figure}
  \begin{center}
    \epsfig{file=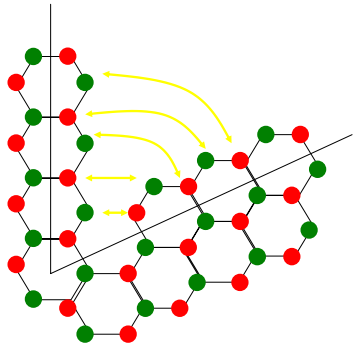,height=7cm}
    \caption{Description of a fivefold ring in the honeycomb structure
    in terms of a disclination. The identification of sites at the two edges
    imply a correspondence between sites from one sublattice and the other.}
    \label{disclination}
\end{center}
\end{figure}

A schematic view of the correspondence of a fivefold ring and a
disclination is shown in Fig.[\ref{disclination}].
The previous analysis shows that the effects on the electronic
states of odd numbered rings in the honeycomb lattice are
approximately described in terms of a gauge field which decays
inversely proportional to the distance to the defect. This scheme
allows us to calculate analytically the low energy electronic
spectrum of closed structures, like C$_{60}$ and higher
fullerenes\cite{GGV92}. The comparison with more detailed
calculations is quite reasonable, and, as expected, improves as
the radius of the system becomes larger.
\subsection{Dislocations.}
The effects induced far away from the core of a dislocation can be
approximated by assuming that the dislocation is made up of two
disclinations of opposite sign. The general model of a
disclination has been given in the previous section, and it can be
directly applied to the case of a dislocation. Its effect on the
low energy electronic spectrum can be approximated by the gauge
field induced by to opposite magnetic fluxes separated by a
distance of the order of the Burgers vector of the dislocation.
This field decays like the inverse of the square of the distance
to the defect.
\subsection{Edge states.}
Tight binding models have shown that, in the vicinity of the edges
of graphene planes, localized states at zero energy can
exist\cite{WS00,W01}. These states are well described by the Dirac
equation used here. The existence of a state at zero energy
implies the existence of a localized wavefunction $( \Psi_1 ( {\bf
\vec{r}} ) , \Psi_2 ( {\bf \vec{r} })$ such that:
\begin{eqnarray}
\left( \partial_x + i \partial_y \right) \Psi_1 ( {\bf \vec{r}} )
&= &0 \nonumber
\\ \left( \partial_x - i \partial_y \right) \Psi_2 ( {\bf \vec{r}} ) &= &0
\end{eqnarray}
These equations are satisfied if $\Psi_1 ( {\bf \vec{r}} )$ is an
analytic function of $z = x + i y$ and $\Psi_2 ( {\bf \vec{r}} ) =
0$, or if $\Psi_1 ( {\bf \vec{r}} ) = 0$ and $\Psi_2 ( {\bf
\vec{r}} )$ is an analytic function of $\bar{z} = x - i y$. We now
consider a semiinfinite honeycomb lattice with an edge at $y=0$
and which occupies the half plane $x>0$. A possible solution which
decays as $x \rightarrow \infty$ is $\Psi_1 ( x , y ) \propto e^{-
k z} = e^{i k y} e^{- k x} , \Psi_2 ( {\bf \vec{r}} ) = 0$. These
solutions satisfy the boundary conditions at $y=0$ if the last
column of carbon atoms belong to the sublattice where the
component $\Psi_1$ is defined. Then, the next column belongs to
the other sublattice, where the amplitude of the state is, by
construction, zero.
\subsection{Vacancies.}
\begin{figure}
  \begin{center}
    \epsfig{file=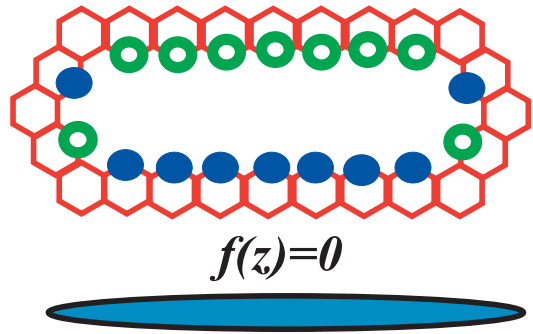,height=5cm}
    \caption{Elongated crack in the honeycomb structure. The crack is such
    that the sites in the upper edge belong to one sublattice, while those
    at the lower edge belong to the other. Bottom: approximate cut in the
    complex plane which can be used to represent this crack at long distances.}
    \label{crack}
\end{center}
\end{figure}
The analysis in the previous section of edge states can be
extended to the existence of localized states near extended
vacancies in the honeycomb lattice. The only possible localized
states can exist at zero energy, where the density of extended
states vanishes. Then, the wavefunctions obtained from the Dirac
equations must be normalizable and analytic on the variables $z =
x+iy$ or $\bar{z} = x-iy$. Extended vacancies with approximate
circular shape can support solutions of the type $\Psi ( ( {\bf
\vec{r}} ) \propto z^{-n} , n > 1$. By using conformal mapping
techniques, solutions can be found with the  boundary conditions
appropriate to the shape of different defect.

A simple case is the elongated crack depicted in
Fig.[\ref{crack}]. A localized solution is described by an
analytic function $f ( z )$ such that ${\rm Re} f ( z ) = 0$ at
the edges of the crack. A family of functions for a crack of half
length $L$, which satisfy these requirements are:
\begin{equation}
f ( z ) = \frac{1}{( z^2 - L^2 )^{n+1/2}}
\end{equation}
\subsection{Random distribution of defects.}
As discussed above, many classes of lattice defects can be
described by gauge fields coupled to the two dimensional Dirac
equation. A random distribution of defects leads to a random gauge
field, with variance related to the type of defect and its
concentration. There is an extensive literature on the problem, as
the model is also relevant to Fractional Quantum Hall states and
to disorder in d-wave superconductors. A random field, when
treated perturbatively, leads to corrections to the wavefunction
renormalization which depend logarithmically in the electronic
bandwidth, in the same manner as the corrections induced by the
long range Coulomb interaction. Hence, disorder is a marginal
perturbation in the Renormalization Group sense, and can be
analyzed using the same approach employed in the study of the
Coulomb interactions.

 Disorder in systems with energy gaps tends
to induce localized states inside the gap. The honeycomb lattice
has a semimetallic density of states. A random field enhances the
density of states at low energies, although the system preserves
its semimetallic character. The density of states at low energies
is changed from $D ( \omega ) \propto | \omega |$ to $D ( \omega )
\propto | \omega |^{1-\delta}$, where $\delta$ depends on the type
of disorder\cite{CMW96,Cetal97,HD02}.
\section{Combined effects of
disorder and the electronic interactions.}
\subsection{The long
range Coulomb interaction.}
The analysis of the Renormalization Group results presented
previously led to the conclusion that the pure graphene system
at low energies is an anomalous Fermi liquid with no short
range interactions.
Inclusion of disorder modelled as
random gauge fields modifies the flow of the couplings
and gives rise to new phases with different physical
properties.
 Similar problems have been considered in relation to
transitions between Fractional Quantum Hall states\cite{Y99}.
There the different types of "extended" disorder are
associated to different gauge couplings that can be treated
with the Renormalization Group technique together with the long range
Coulomb interaction.

\begin{figure}
  \begin{center}
    \epsfig{file=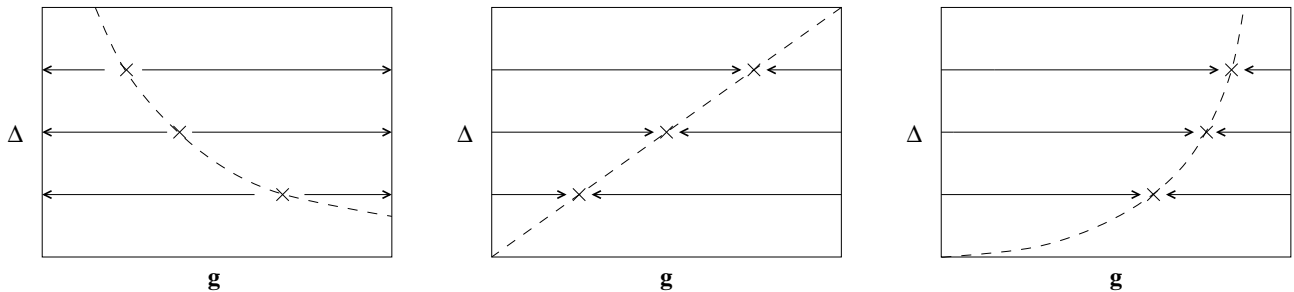,height=3.8cm}
    \caption{Phase diagram calculated by renormalizing the strength of
    the disorder, $\Delta$, and the Coulomb interaction, $g$, at the same time. The three
    plots correspond to three types of disorder. Left: random on site potential.
    Center: Random disorder in the lattice structure. Right: Random correction to
    the hoppings.}
    \label{Phasediagram}
\end{center}
\end{figure}

The values of the effective coupling constants at low energies
are modified by the new interactions giving rise to a rich phase
diagram with new
phases with different physics depending of the type of disorder
%allow us to identify the possible phases, as function of the
%strength of the disorder and the Coulomb interaction.
A schematic
plot of the flow obtained for different types of disorder is shown
in Fig.[\ref{Phasediagram}]. The most interesting
phase is the one induced by the presence of
random disorder in the lattice structure(center).
It shows the existence of new
phases, where the effect of the Coulomb interaction, which tends
to lower the density of states at the Fermi energy \cite{pil1}, and the
disorder, which has the opposite tendency, balance each other.
This phase is stable, within the limitations of the calculation
presented here, for certain types of disorder\cite{SGH03}.
\subsection{Short range interactions.}
As mentioned earlier, short range interactions, such an onsite
Hubbard term $U$ are irrelevant, in the sense that their effect
can be analyzed within standard perturbation theory without
encountering divergences. This is due to the vanishing density of
states at the Fermi level. The Hubbard model at half filling, in a
two dimensional square lattice leads to a highly singular
perturbation expansion, due to the diverging density of states at
the Fermi level. As mentioned above, the density of states at low
energies is increased by the presence of disorder. This, in turn,
enhances the effect of short range interactions.

Short range interactions can lead to a variety of phases at low
temperatures. In the absence of disorder, an onsite Hubbard term
favors antiferromagnetism. An antiferromagnetic phase, however, is
likely to be suppressed by disorder, especially by the presence of
odd numbered rings in the lattice. Then, the next leading
instability that such an interaction can induce is towards a
ferromagnetic phase.

If a magnetic phase does not appear, electron electron
interactions, even when they are repulsive, will lead to an
anisotropic ground state. The existence of two inequivalent Fermi
points in the Brillouin zone suggests that the superconducting
order parameter induced by a repulsive interaction will have
opposite sign at each point. The corresponding symmetry is p-wave.
No that disorder, in addition to the enhancement of the density of
states mentioned already, will lead to pair breaking effects in an
anisotropic superconducting phase.
\subsection{Interactions between localized states.}
As mentioned in the previous section, vacancies and cracks in the
honeycomb lattice induce localized states at the Fermi energy.
These states will become polarized in the presence of repulsive
interactions, as this polarization implies no cost of kinetic
energy. Then, we can expect that lattice defects will nucleate
magnetic moments in their vicinity. These moments can be large, as
the number of localized states is proportional to the number of
sites at the perimeter of the defect. Note that this mechanism is
intrinsic to the graphene structure, and it does not require the
trapping of magnetic ions near the defects.

The moments near different defects polarize the conduction band of
the surrounding medium, leading to an effective RKKY interaction.
In an ordinary metal this interaction is made up of an oscillatory
and a decaying term as function of distance, and it can be of
either sign, leading to frustration and spin glass effects. The
graphene plane considered here, however, does not have a Fermi
surface, so that the induced RKKY interaction does not oscillate.
A simple analysis, using the analytical expression for the
susceptibility discussed in earlier chapters, gives:
\begin{equation}
J_{RKKY} ( {\bf \vec{r}} ) \sim U^2  \int d^2 {\bf k} e^{i {\bf
\vec{k}} {\bf \vec{r}} } \chi ( {\bf \vec{k}} ) \sim U^2
\frac{a^4}{v_F | {\bf \vec{r}} |^3} \label{RKKY}
\end{equation}
where $U$ is the magnitude of the onsite Hubbard term. Hence, the
RKKY interaction is ferromagnetic, and it decays as $r^{-3}$ as
function of the distance between local moments.
\section{Coupling between graphene layers.}
\subsection{Coulomb interactions.}
So far, we have considered the properties of isolated graphene
layers. As the source of most of the unusual properties reported
here is the long ranged Coulomb interaction, it is important to
consider the screening effects of neighboring layers.

The calculation of the full dielectric constant of a set of
metallic or semimetallic layers in terms of single layer
properties can be done analytically. The dielectric function of
the system can be written as\cite{GGV01}:
\begin{equation}
\frac{1}{\epsilon ( {\bf \vec{q}} , \omega )} = \frac{\sinh ( |
{\bf \vec{q}} | d )}{\sqrt{\left[\cosh ( | {\bf \vec{q}}| d ) + (
2 \pi e^2 / | {\bf \vec{q}} ) | \sinh ( | {\bf \vec{q}} | d )
\chi_0 ( {\bf \vec{q}} , \omega ) \right]^2 - 1}}
\end{equation}
where $\chi_0 ( {\bf \vec{q}} , \omega )$ is the charge response
function of an isolated layer, and $d$ is the interlayer distance.
This response function is finite for $|{\bf \vec{q}}| \ll d^{-1}$,
so that the interactions remain long range.
\subsection{Interlayer hopping.}
The electron--electron interaction modifies the quasiparticle
propagator, as discussed above. The electrons within the layers
are dressed by a cloud of virtual excitations. This cloud cannot
follow an electron which hops between neighboring layers, reducing
the effective tunnelling element.

In conventional Fermi liquids, this renormalization of the
interlayer hopping is finite, and it can be calculated using
perturbation theory. In the model studied here, this calculation
leads to divergencies, and it resembles closely the analysis of
the electron self energy sketched previously.

The ^^ ^^ orthogonality catastrophe " which results from the
virtual excitation of electron-hole pairs has been extensively
discussed in connection to the physics of mesoscopic
systems\cite{Betal00}, and it has also been applied to the related
problem of tunnelling between two dirty metallic
layers\cite{TL01}. Similar procedures can be used in the present
case. The interlayer hopping acquires a multiplicative
renormalization which makes it vanish at low energies, even in
clean samples. This calculation is consistent with the extreme
anisotropy observed in some experiments\cite{Oetal03}.
\section{Conclusions.}
We have discussed a simplified model for the long wavelength
electronic properties of graphene planes. The interplay between
the semimetallic properties of the planes and the long range
interactions leads to the existence of a variety of interesting
effects:
\begin{itemize}
\item The model, in the absence of disorder, shows deviations from
Landau's theory of a Fermi liquid. The quasiparticles are strongly
renormalized, and their lifetimes do not follow the usual $\Gamma
( \epsilon ) \propto ( \epsilon - \epsilon_F )^2$ behavior. The
low energy electronic properties of the system can be thought as
similar to the ^^ ^^ pseudogap " regime in the superconducting
cuprates.  \item Disorder can be incorporated into the model in a
simple way. While the interactions tend to deplete the electronic
density of states near the Fermi energy, disorder leads to its
enhancement. The resulting competition induces the existence of an
^^ ^^ incoherent metal " regime at low energies, similar to the
one dimensional Luttinger liquid, although stabilized by the
disorder. \item Large lattice deformations can nucleate localized
electronic states in their vicinity\footnote{Note that these
states are built up from the $\pi$ orbitals of each layer, and the
$\sigma$ bonds remain unaltered.}. These states can lead to the
formation of local moments. The absence of a finite Fermi
wavevector implies that the RKKY interaction mediated by the
conduction electrons does not change sign, and it is
ferromagnetic. Hence, the frustration which leads to spin glass
behavior in metals with magnetic impurities is absent in this

case. \item The screening cloud around quasiparticles suppresses
interlayer tunnelling, enhancing the anisotropy of the electronic
properties.
\end{itemize}
%Funding from MCyT (Spain) through grant MAT2002-0495-C02-01 is
%acknowledged.
%\bibliography{sns04}

\end{document}